\documentclass[prl,aps,10pt,nofootinbib,twocolumn,
nobibnotes,
preprintnumbers,superscriptaddress,notitlepage]{revtex4-2}

\usepackage{xcolor}
\usepackage{graphicx}  
\usepackage{amsfonts}  
\usepackage{amsmath}
\usepackage{amssymb} 
\usepackage{adjustbox}
\usepackage{hyperref}
\makeatletter
\def\l@subsubsection#1#2{} 
\makeatother
\hypersetup{
   pdftitle={Dark GFT},
   colorlinks=true,
   linkcolor=blue,
   citecolor=green,
   urlcolor=blue
   }

\def\dd{{\rm d}}
\DeclareMathOperator{\sgn}{sgn}

\begin{document}

\title{Emergent dark sector in group field theory cosmology}
\author{Andrea Calcinari}
\email{andrcalc@ucm.es}
\affiliation{Departamento de F\'isica Te\'orica and IPARCOS, Facultad de Ciencias F\'isicas, Universidad Complutense de Madrid, Plaza de Ciencias 1, 28040 Madrid, Spain}
\author{Federico Greco}
\email{federico.greco.1@phd.unipd.it}
\affiliation{Dipartimento di Fisica e Astronomia “G. Galilei”, Università degli Studi di Padova, via Marzolo 8, I-35131 Padova, Italy}
\affiliation{INFN, Sezione di Padova, via Marzolo 8, I-35131 Padova, Italy}
\author{Daniele Oriti}
\email{doriti@ucm.es}
\affiliation{Departamento de F\'isica Te\'orica and IPARCOS, Facultad de Ciencias F\'isicas, Universidad Complutense de Madrid, Plaza de Ciencias 1, 28040 Madrid, Spain}
\author{Pietro Pellecchia}
\email{pietro.pellecchia2@unina.it}
\affiliation{Dipartimento di Fisica Ettore Pancini, Università di Napoli “Federico II”,
Complesso Univ.\ Monte S.\ Angelo, I-80126 Napoli, Italy}
\affiliation{INFN, Sezione di Napoli, Complesso Univ.\ Monte S.\ Angelo, I-80126 Napoli, Italy}

\begin{abstract}

We develop an analytical treatment of the emergent cosmological dynamics induced by local polynomial interactions in a deparametrised group field theory model, moving beyond the usual non-interacting approximation. For a single field mode with quartic and sextic couplings we obtain a closed-form generalised Friedmann equation within a controlled Gaussian regime. The dynamics preserves the quantum bounce while generating effective dark matter and dark energy contributions as collective quantum-geometric phenomena, 
providing the first derivation of both components from a single model. Matching the resulting dark-energy-to-dark-matter density ratio to observations places concrete phenomenological constraints on the fundamental theory, linking the cosmic coincidence problem directly to the underlying quantum gravity dynamics. Our results extend to arbitrary even polynomial interactions, establishing a systematic dictionary between microscopic interaction orders and effective equations of state.

 \end{abstract}

\maketitle

Cosmology confronts quantum gravity with challenges at both ends of cosmic history. At early times, the classical Big Bang singularity signals the breakdown of general relativity where quantum-gravitational corrections become important~\cite{Hawking:1970zq}, and possibly the breakdown of the continuum picture of spacetime itself. At late times, the standard cosmological model accounts for cosmic acceleration and structure formation only by positing dark energy (as a cosmological constant $\Lambda$) and cold dark matter (CDM) as phenomenological ingredients \cite{weinberg2008cosmology}. The \textit{dark sector} of the $\Lambda$CDM model, however, lacks any derivation from an underlying fundamental theory~\cite{Peebles:2002gy,Bertone:2016nfn}. A successful quantum gravity theory should resolve the singularity and could explain how dark matter and dark energy emerge from the same microscopic dynamics.

Among the background-independent approaches built on discrete structures, Group Field Theory (GFT) stands out for treating spacetime geometry itself as emergent from a many-body collective dynamics of such structures~\cite{GFTquantumST_Oriti,ORITI2017235,Bogo}. Its fundamental degrees of freedom are described by fields on group manifolds \cite{FreidelGFT, oriti_microscopic_2012}, and their quanta are naturally interpreted as building blocks of quantum geometry, closely related to spin-network states of loop quantum gravity~\cite{Oriti_GFT2ndLQG}. Numerous works have shown that the coarse-grained dynamics of GFT models recovers effective Friedmann dynamics at large volumes and replaces the Big Bang singularity with a quantum bounce at small volumes~\cite{Oriti_2016,BOriti_2017,relham_Wilson_Ewing_2019,Gielen_2020}. While these results are robust, surviving across different models and quantisation schemes~\cite{Marchetti2021,PWGFT,Marchetti:2024nnk}, they mostly rely on the \textit{free} theory. 

More recently, after early hints of a dust-like term at quartic order~\cite{deCesare:2016rsf,Gielen_2020}, efforts to incorporate interactions have shown that sixth-order couplings can reproduce dynamical dark energy scaling. This is obtained within mean-field dynamics on specific coherent states~\cite{Marchetti2021} and with simplified interaction kernels either modelled phenomenologically ~\cite{Oriti:2021rvm,Pang:2025jtk} or subject to structural restrictions \cite{Ladstatter:2025kgu} and treated numerically \cite{Marchetti:2025jze}. Moreover, dark energy is read off asymptotically, in a regime of cosmic volume beyond analytic control. Therefore, the robustness of the result remains an open issue; no emergent dark matter has been clearly established, and the two dark components have never been obtained together.

In this Letter we bridge this gap. We study a GFT model for a single mode with both quartic and sextic interactions and derive the effective cosmological equations analytically, in a relational deparametrised setting, from the quantum operator equations using a Gaussian closure. The closure is algebraic and holds for arbitrary Gaussian states, yielding an autonomous Friedmann-like equation in the dynamically invariant zero-displacement sector, and generalises in a systematic way to interactions of any even order. The resulting dynamics features a primordial quantum bounce together with terms corresponding to both the dark matter and dark energy components, arising as collective effects of the underlying many-body quantum geometry. Importantly, at any given volume the relative weight of the two contributions is fixed by the fundamental theory alone, and can be matched to the observed dark-energy-to-dark-matter ratio to obtain a direct phenomenological constraint on the microscopic parameters.

{\bf Group field theory.} --- GFTs extend random matrix and tensor models \cite{DiFrancesco:1993cyw,Gurau:2011xp} with additional Lie group-theoretic data featuring also in loop quantum gravity and spin foam models~\cite{DePietri:1999bx,Oriti_GFTandLQG}, with a quantum geometric interpretation. In particular, the perturbative expansion of the GFT partition function generates a sum over cellular complexes, with Feynman amplitudes corresponding to the discrete spacetime histories of spin foam models~\cite{Reisenberger_2000} and lattice gravity path integrals \cite{Baratin:2011hp,Baratin:2011tx}. For the simplicial models used in cosmological applications, the fundamental variable is a (here, real) field having as arguments $g_I:=(g_1,g_2,g_3,g_4)\in{\rm SU}(2)^4$, matching the structure of four-valent spin network nodes and 3-simplices~\cite{Oriti_GFT2ndLQG}, and a real variable $\chi$ acting (at the level of GFT Feynman amplitudes) as a minimally coupled massless scalar field:
\begin{equation}\label{eq_field}
    \varphi: {\rm SU}(2)^4 \times \mathbb{R} \rightarrow \mathbb{R}\,,
\end{equation}
with $ \varphi(g_I,\chi) = \varphi(g_I h,\chi)$ for all $h \in \rm SU (2)$.

Decomposing the field into Peter--Weyl modes $\varphi_J$, where $\pm J = (j_I, \pm m_I, \iota)$ encodes irreducible representations $j_I \in \mathbb{N}/2$, magnetic indices $m_I \in [-j_I,j_I]$, and the intertwiner label $\iota$, the GFT action takes the form
\begin{equation}\label{eq_action}
    S = \frac{1}{2}\int \mathrm{d}\chi \sum_{J} \varphi_{J}(\chi) \Big(K^{(0)}_{J}+K^{(2)}_{J} \partial_\chi^2\Big) \varphi_{-J}(\chi) + U[\varphi]\,,
\end{equation}
where we assume the minimal form of the kinetic term respecting the dynamical symmetries of a minimally coupled free massless scalar field~\cite{Gielen_2016,Marchetti:2022nrf,Carrozza:2016vsq}, while $U[\varphi]$ encodes generically non-local interactions on the group manifold.

We follow a canonical quantisation \cite{Gielen:2024sxs}---well suited to the cosmological sector \cite{GFTcosmoLONGpaper}---and deparametrise classically by choosing $\chi$ as the physical clock \cite{relham_Wilson_Ewing_2019} (see~\cite{PWGFT,Marchetti2021,Marchetti:2024nnk} for relational clocks selected at the quantum level). Defining the conjugate momenta $\pi_J := -K_J^{(2)} \partial_\chi \varphi_{-J}$ and the corresponding \textit{relational} Hamiltonian $H$, one promotes $\varphi_J$ and $\pi_J$ to operators with $ \left[\hat \varphi_J(\chi)\,,\,\hat\pi_{J'}(\chi) \right]={\rm i}\delta_{J,J'}$. Introducing the corresponding ladder operators with canonical commutation relations $[\hat a_J , \hat a^\dagger _{J^\prime}]= \delta_{J J^\prime}$ allows one to build the physical Hilbert space of the theory as a Fock space over the no-geometry state, $\hat{a}_J|0 \rangle=0$, whose one-particle excitations $	|\, \vcenter{\hbox{\adjincludegraphics[width=.023\textwidth]{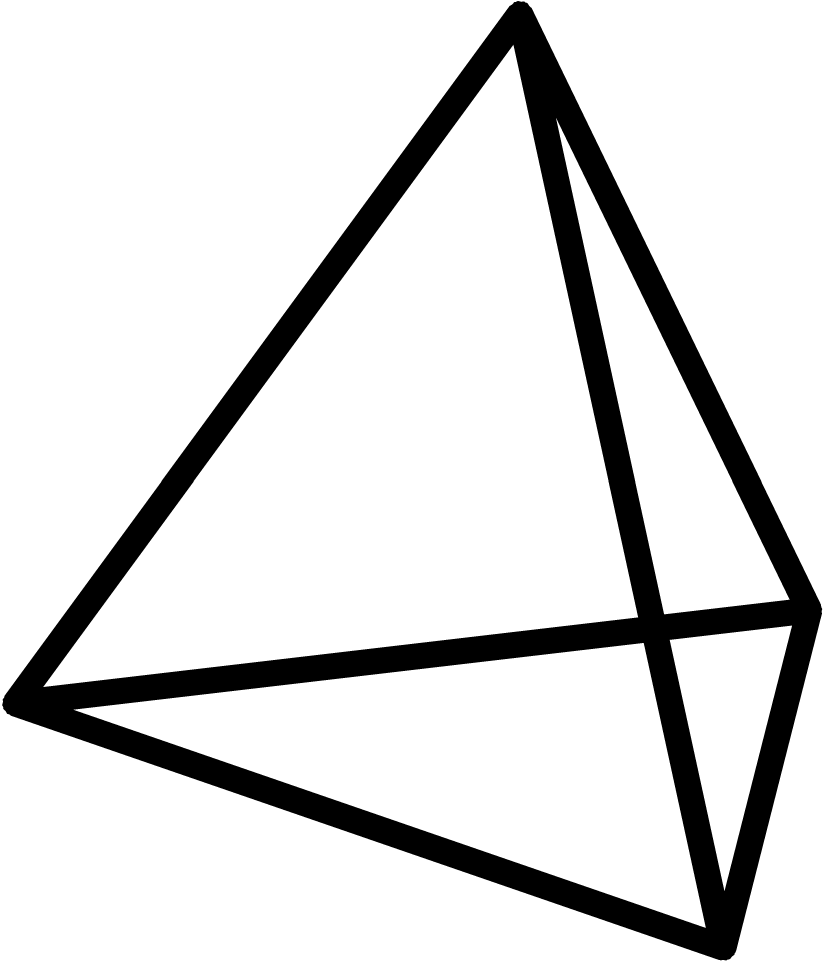} }}\rangle= \hat{a}^\dagger_J |0\rangle$ represent quantum tetrahedra endowed with group-theoretic data $J$ \cite{relham_Wilson_Ewing_2019,Gielen_2020}. On such a Fock space one defines the number and volume operators for every $J$ as
\begin{equation}\label{number operator}
\hat N_J := \hat a^\dagger_J \hat a_J\,,\qquad \hat V_J := v_{J}\hat N_J \,,
\end{equation}
where $v_J$ represents the volume carried by a quantum of geometry associated to the mode $J$ \cite{Baez_1999,Bianchi:2010gc}.

The total spatial volume $\hat V (\chi) := \sum_{J} \hat V_{J}(\chi)$, built from the individual contributions of quantum tetrahedra, is then the macroscopic observable of cosmological interest, and its relational evolution is governed by the Heisenberg equations of motion generated by the quantum Hamiltonian $\hat{H}$.

{\bf Emergent bouncing cosmology from free GFT.} --- In the free theory ($U[\varphi]=0$), the sign of $K_{J}^{(0)}K_{J}^{(2)}$ dictates whether each mode Hamiltonian describes a harmonic oscillator ($>0$) or single-mode squeezing ($<0$) in a suitable basis \cite{PWGFT}, with rate $ \omega_J:= -\sgn(K^{(0)}_{J})\,{|K^{(0)}_{J}/K^{(2)}_{J}|}^{1/2} $. The squeezing Hamiltonian generates GFT quanta at an exponential rate, rapidly dominating the conserved number of particles in the oscillatory modes \cite{toy,relham_Wilson_Ewing_2019,relhamadd}. Importantly, the squeezed mode with the largest $|\omega_J|$ grows exponentially faster than all others, ultimately driving the dynamics. The resulting condensation---in the sense of a macroscopic occupation of a single $J$---emerges dynamically from the free theory, rather than being imposed on the state, yielding an effective single-mode regime at late relational times~\cite{Gielen_lowspin}. Such a production of quanta drives an exponential growth of the volume of the emergent universe, with the squeezing rate $\omega_J$ translating directly into the expansion rate.

This justifies restricting the analysis to the single-mode theory with Hamiltonian 
\begin{equation}\label{free hamiltonian}
    \hat{H}= -\frac{\omega}{2} (\hat{a}^2+\hat{a}^{\dagger 2}) \,,
\end{equation}
where the index $J$ is dropped hereafter. Such an approximation accurately describes a broad class of GFT models (especially at late relational times) and suffices to recover effective cosmological dynamics compatible with classical general relativity~\cite{Oriti_2016,BOriti_2017,relham_Wilson_Ewing_2019,Gielen_2020}. Specifically, the evolution of the macroscopic volume operator $\hat{V}(\chi) = v  \hat{N}(\chi)$ follows directly from the algebraic structure of the system. Because $\hat H$ belongs to the $\mathfrak{su}(1,1)$ algebra spanned by $\hat a^2$, $\hat a^{\dagger2}$ and $\hat N$ \cite{Gielen_2020,Gauss}, the Heisenberg equations of these generators close among themselves with constant coefficients, and the resulting linear system is solvable in closed form. This yields the explicit relational evolution of the expectation values
\begin{equation}\label{def n e k}
    n(\chi):= \langle  \hat{a}^\dagger(\chi)\,\hat{a}(\chi)\rangle 
    \,, \qquad k(\chi):=\langle \hat a(\chi)^2\rangle \, ,
\end{equation}
allowing us to cast the dynamics of $V(\chi) := v  \,n(\chi)$ as
\begin{equation}\label{Calcinari}
   \left(\frac{V'(\chi)}{V(\chi)}\right)^2 = 4\omega^2 \left(1+ \frac{v}{V(\chi)}  - \frac{v^2\mathcal{I}_0}{V(\chi)^2} \right)\,,
\end{equation}
with $\mathcal{I}_0:= n_{0}(n_{0}+1)-(\mathrm{Im}\,k_{0})^{2}$ encoding the initial conditions $k_0:= k(\chi_0)$ and $n_0:= n(\chi_0)$. We emphasise that \eqref{Calcinari} is exact and holds for arbitrary states since the $\mathfrak{su}(1,1)$ closure is a statement at the operator level. It is precisely this closure that interactions break.

At late relational times, or large volumes $V(\chi)\gg v$, \eqref{Calcinari} reduces to the classical relational Friedmann equation for a flat FLRW universe where a minimally coupled massless scalar field $\chi$ is used as a clock, $(V'/V)^2=12\pi G$. Importantly, this allows the identification
\begin{equation}\label{Newton bare}
     G=\frac{\omega^2}{3\pi} \,,
\end{equation}
showing the emergence of Newton’s constant from the couplings of the fundamental theory. Conversely, at early times (small volumes), the terms proportional to $V^{-1}$ and $V^{-2}$ modify the classical trajectory, generically preventing the volume from vanishing (since physical states satisfy $\mathcal{I}_0\geq 0$ by the Cauchy--Schwarz inequality), thereby replacing the Big Bang singularity with a quantum bounce~\cite{Oriti_2016,BOriti_2017,relham_Wilson_Ewing_2019,Gielen_2020}.

{\bf Interacting theory and emergent dark sector.} --- While \eqref{Calcinari} provides a robust description of the early-universe bounce and of the subsequent classical dynamics, non-linear interactions inevitably become important at late relational times, requiring an extension
beyond the free theory. Although GFT interactions in general induce inter-mode couplings, choosing a kernel diagonal in the representation labels isolates the single-mode dynamics, and this arises naturally in isotropic simplicial models~\cite{Oriti_2016,Pang:2025jtk}. Specifically, we consider $\hat\varphi^4$ and $\hat\varphi^6$ interactions, where  $\hat\varphi \propto \hat a + \hat a^\dagger$ and we absorb the proportionality constant into the couplings. We thus generalise the models of \cite{Gielen_2020,Gielen:2023han} with the single-mode Hamiltonian
\begin{equation}\label{Interacting_Hamiltonian}
       \hat  H= -\frac{\omega}{2} (\hat{a}^2+\hat{a}^{\dagger 2}) -\frac{g\omega}{4} (\hat{a}+\hat{a}^\dagger)^4-\frac{\lambda\omega}{6} (\hat{a}+\hat{a}^\dagger)^6 \,,
\end{equation}
with $g, \lambda > 0$ dimensionless couplings, assumed positive so that the interactions share the sign of the kinetic term.

Paralleling the free-theory derivation of \eqref{Calcinari}, we now obtain the effective Friedmann dynamics generated by the interacting Hamiltonian. The main steps are outlined here and the full derivation is given in the Appendix. 

The higher-order terms in \eqref{Interacting_Hamiltonian} explicitly break the $\mathfrak{su}(1,1)$ operator closure of the free theory. As a result, the evolution of $n(\chi)$---and consequently of the volume $V(\chi)= v\, n(\chi)$---couples to higher-order correlation functions, generating an infinite tower of coupled equations for the moments. To truncate this infinite hierarchy we impose a Gaussian closure, namely we assume the state remains approximately Gaussian so that Wick's theorem reduces any correlator to first and second moments. Since the free Hamiltonian preserves Gaussianity, this approximation is reliable as long as the interaction terms remain subdominant to the kinetic term. Relative to the quadratic term, the interactions in \eqref{Interacting_Hamiltonian} have expectation values scaling as $gn$ and $\lambda n^{2}$, so the closure requires 
\begin{equation}\label{gnl}
     g n\ll1\,, \qquad   \lambda  n^2\ll1 \,.
\end{equation}
This makes the analysis reliable at least 
up to $ V\ll \min\{g^{-1},\lambda^{-1/2}\}\,v
$, beyond which the approximation is expected to break down.

The Gaussian truncation yields a closed system of five real equations---as many as the parameters of a Gaussian state---for $n$ and the complex functions $k$ and $\alpha:=\langle\hat a\rangle$---see \eqref{eq:SM_five}. The system admits two constants of motion: the Hamiltonian expectation value $\langle\hat H\rangle$ (conserved exactly since \eqref{Interacting_Hamiltonian} is time independent), and the symplectic invariant of the covariance matrix
\begin{equation}
    Q:=\big(n-|\alpha|^2+\tfrac{1}{2}\big)^2-\big| k-\alpha^2\big|^2\,,
\end{equation}
conserved strictly within the Gaussian approximation. 

In the free theory the expansion rate depends on the state only through ${\rm Im}\,k$, which the $\mathfrak{su}(1,1)$ closure fixes at any given volume; \eqref{Calcinari} holds in any state because the \emph{uncentred} invariant
$(n+\tfrac12)^{2}-|k|^{2}$ is conserved. The interacting dynamics, on the other hand, couples the expansion rate separately to the mean momentum of the state (i.e., ${\rm Im}\, \alpha$) and to its field-momentum correlation (i.e., ${\rm Im}(k-\alpha^2)$), while the invariants constrain only a single combination of the two (see \eqref{eq:SM_fibre}). Thus, at a given volume, the conserved quantities do not determine the expansion rate, which ranges from zero up to a maximum (see \eqref{eq:SM_angle}). Consequently, \eqref{Calcinari} admits no closed generalisation in the interacting theory for displaced states. This is not a limitation of our method: while the full system can be integrated numerically, no single Friedmann equation determined by $V$ together with $\langle\hat H\rangle$ and $Q$ exists for arbitrary displaced states.

We therefore restrict to the zero-displacement sector ($\alpha = 0$), as realised by squeezed thermal states. This is preserved exactly by the dynamics, since $\hat H$ contains only even powers of $\hat a+\hat a^{\dagger}$, and the system closes on three real equations. In particular, the equation for $n$ simplifies to $n'=2\omega\,{\rm Im}\,k\,(1+3gS+15\lambda S^{2})$ with $S:=2{\rm Re}\,k+2n+1$, the bracketed factor measuring the enhancement due to interactions. This equation can be decoupled algebraically using the two constants of motion, where now $Q=(n+1/2)^{2}-|k|^{2}$: their conservation laws eliminate $k$ in favour of $n$. Specifically, $Q$ fixes $({\rm Im}\,k)^{2}$, while $\langle\hat H\rangle$ determines $S$ as the positive root of a cubic~\eqref{eq:SM_cubic}. The result is an autonomous equation fixed by the initial data $(n_{0},k_{0})$ which, via $V=v\,n$, yields a modified effective Friedmann equation:
\begin{equation}\label{eq:ExactFE}
\begin{aligned}
    \left(\frac{V'}{V}\right)^{2}&=4\omega^{2}\big(1+3gS+15\lambda S^{2}\big)^{2}
    \left(1+\frac{v}{V}-\mathcal{I}\,\frac{v^{2}}{V^{2}}\right)\,,    
\end{aligned}
\end{equation}
where $\mathcal{I}:=Q-\frac14+(\mathrm{Re}\,k)^{2}$. Similarly to the free-theory setting, $\mathcal{I}$ is non-negative for physical states since positivity of the covariance matrix requires $Q\geq 1/4$, so the bounce persists generically in the interacting theory. Every quantity on the right-hand side of \eqref{eq:ExactFE} is a function of $V$ alone, fixed by the initial data. We display it in terms of $S=S(V)$, whose closed form is given in \eqref{eq:SM_cardano} in the Appendix, because its general structure is far more transparent than the fully explicit expression.

The Friedmann equation \eqref{eq:ExactFE} provides a clear generalisation of \eqref{Calcinari} within the zero-displacement Gaussian family: setting $g=\lambda=0$ reduces the cubic to a linear equation, so that ${\rm Re}\,k={\rm Re}\,k_{0}$ and $\mathcal{I}$ naturally reduces to $\mathcal{I}_{0}$. Moreover, for $\lambda=0$, \eqref{eq:ExactFE} is the quantum counterpart of the quartic model of Ref.~\cite{Gielen_2020}, where an exact equation was obtained for the corresponding classical system, and the quantum treatment was only perturbative in the coupling and evaluated on particular pure states. By contrast, \eqref{eq:ExactFE} is exact in the couplings and holds for every zero-mean Gaussian state, mixed ones included.

Although the dynamics is available in closed form, we do not extrapolate it beyond the regime of validity and expand \eqref{eq:ExactFE} in the couplings to expose the cosmological implications of interactions. This comes at no loss of control since the expansion is governed by the small parameters $gn$ and $\lambda n^{2}$, which by \eqref{gnl} require in particular both couplings to be small, and retaining higher orders would not improve the accuracy. To first order one finds:
\begin{widetext}
\begin{equation}\label{Taylor}
         \left(\frac{V'}{V}\right)^{2}=4\omega^{2}\Big[(1+g A_{0}+\lambda B_{0})
         +\frac{v}{V}(1+gC_{0}+\lambda D_{0})
         -\frac{v^2}{V^{2}}(\mathcal{I}_{0}+gE_{0}+\lambda F_{0})
         +\frac{(12g+\lambda G_{0})}{v} V+\frac{120\lambda}{v^2} V^{2}
         \Big]\,,
\end{equation}
\end{widetext}
where the coefficients $A_0,\dots ,G_0$ depend solely on initial conditions and are given in \eqref{eq:SM_taylorcoeff}, and the neglected terms are second order in $gn$ and $\lambda n^{2}$. By \eqref{gnl}, the coupling-dependent terms remain small throughout the regime of validity: interactions correct the \eqref{Calcinari} without dominating.

The expansion \eqref{Taylor} reveals that GFT interactions alter the cosmic evolution through two distinct mechanisms. First, they renormalise the free-theory coefficients of \eqref{Calcinari}, preserving the bounce and dressing the bare gravitational coupling~\eqref{Newton bare} into an effective Newton's constant:
\begin{equation}\label{Newton full}
     G = \frac{\omega^2}{3\pi}\left(1+gA_0 +\lambda B_0 \right)\,.
\end{equation}
Note that the dressing is state-dependent as $A_{0}$ and $B_{0}$ involve initial data; since the couplings are small, it is nonetheless a small correction and $G$ stays positive.

Second, the interactions generate entirely new dynamical terms: the quartic coupling $g$ drives a term linear in the volume, while the sextic coupling $\lambda$ drives a quadratic one, and additionally dresses the linear one through $\lambda G_{0}$. These new terms are instances of a general pattern: as we show in the appendix, a $\hat{\varphi}^{2\kappa}$ interaction contributes a term $\propto V^{\kappa-1}$ with a coefficient fixed by its own coupling alone, together with dressings of the lower orders. 

These new contributions are precisely the powers carried by dust and by a cosmological constant in the \textit{classical} relational Friedmann equation of general relativity, for a spatially flat universe with a minimally coupled massless scalar as clock~\cite{Gielen_2020,Oriti:2021rvm,Gauss}:
\begin{equation}\label{GR equation}
    \left( \frac{V'}{V} \right)^2=12 \pi G\left[1+\frac{ \rho^0_{\rm M} }{\rho^0_\chi V_0}V+\frac{ \Lambda}{8\pi G\rho^0_\chi V_0^2} V^2\right]\,.
\end{equation}
Here $\Lambda$ is the cosmological constant, while $\rho^0_{\rm M}$, $\rho^0_\chi$, and $V_0$ respectively denote the energy density of a pressureless component, the scalar clock energy density, and the volume at a given reference time. The constant term is the contribution of the clock itself, a stiff fluid with equation of state $w=1$, 
which is what \eqref{Calcinari} already reproduced. Its energy density is fixed by the conserved momentum as $\pi_{\chi}^{2}=2V^{2}\rho_{\chi}=2V_{0}^{2}\rho^{0}_{\chi}$.

While $\rho_{\rm M}^0$ in \eqref{GR equation} may denote any pressureless source, which in $\Lambda$CDM is cold dark matter together with a subdominant baryonic component, our quantum gravity model contains no matter beyond the clock. What the quartic interaction supplies is a collective effect of quantum geometry, not new matter degrees of freedom, which enters \eqref{Taylor} exactly as a pressureless fluid with $w=0$. It is therefore a candidate for the dark matter component alone, since ordinary matter in our formalism corresponds to additional degrees of freedom besides quantum geometry (like the clock $\chi$), and we denote it $\rho^0_{\rm DM}$ from here on. Whether it also clusters and 
reproduces the structure growth characteristic of cold dark matter is a question about inhomogeneous perturbations of the (single-mode) GFT condensate, and lies beyond the scope of the homogeneous sector considered here.

Comparing \eqref{Taylor} and \eqref{GR equation} power by power in $V$, the standard dark components are determined by the fundamental GFT couplings as
\begin{equation}\label{dm e de}
        \Lambda = \frac{ 320\lambda\omega^2 V_0^2\rho^0_\chi}{v^2}\,, \quad\rho^0_{\rm DM} =\frac{(12g+\lambda G_0)V_0\rho^0_\chi }{v\left(1+gA_0 +\lambda B_0 \right)} \,.
\end{equation}
The first systematically reproduces the structure of the effective cosmological constant found from sixth-order interactions in Refs.~\cite{Pang:2025jtk,Marchetti:2025jze}, while the second provides the first closed-form expression of a dark matter component directly in terms of the fundamental GFT couplings---both within the same model. Here $\Lambda>0$ follows from $\lambda>0$, while $\rho^{0}_{\rm DM}>0$ holds naturally provided the initial data satisfy $G_{0}\geq0$, which we adopt throughout. 

In general, a contribution $\propto V^{1-w}$ in \eqref{GR equation} is a perfect fluid of equation of state $w$ (see end of Appendix), so each interaction order introduces a new effective equation-of-state parameter $w_{\kappa}=2-\kappa$, on top of dressing the lower ones (see \cite{deCesare:2016rsf} for related results). The clock corresponds to $\kappa=1$, the quartic and the sextic give dust and a cosmological constant, while an eighth-order interaction would generate a phantom component.

While the expressions in \eqref{dm e de} cannot be disentangled from the clock energy density $\rho^0_\chi$ individually, their ratio can. In the standard $\Lambda$CDM model the dark energy and dark matter density parameters are
\begin{equation}
     \Omega_\Lambda:=\frac{\Lambda}{8\pi G\rho_c} \,,\qquad \Omega_{\rm DM}:=\frac{\rho^{\text{today}}_{\rm DM}}{\rho_c} \,,
\end{equation}
where $\rho_c$ is the present critical density and $\rho^{\text{today}}_{\rm DM}$ is the present dark matter density~\cite{weinberg2008cosmology}, which coincides with $\rho^0_{\rm DM}$ if the reference time in \eqref{GR equation} is taken to be the present one, $V_0=V_{\text{today}}$. Current observations constrain their ratio to $\Omega_\Lambda/\Omega_{\rm DM} \simeq 2.6$~\cite{Planck:2018vyg}. Using \eqref{dm e de} one finds
\begin{equation}\label{pheno}
     \frac{\Omega_\Lambda}{\Omega_{\rm DM}}=\frac{120\lambda }{(12g+\lambda G_0)}\frac{V_{\text{today}}}{v}
=\frac{V_{\text{today}}}{V_{\star}}\,,
\end{equation}
where $V_{\star}:=v(12g+\lambda G_{0})/(120\lambda)$ is the volume at which the two interaction terms of \eqref{Taylor} contribute equally, fixed by the fundamental theory. Remarkably, the clock energy density, $\omega$ and the dressed Newton constant \eqref{Newton full} cancel out, the initial data surviving only through $G_{0}$. A dimensionless cosmological observable thus fixes the present volume of the universe in units of a scale built from the couplings of the fundamental theory, $V_\text{today}\simeq 2.6\, V_\star$. 

Consistency then turns \eqref{pheno} into a constraint on the fundamental theory. The observed ratio places the present volume a few times above $V_{\star}$, so the derivation stays within its regime of validity only if the volume at which the two dark components cross lies far below the volume at which the Gaussian description fails. Using \eqref{gnl} one obtains $g^{2}/\lambda\leq(10\,\Omega_{\rm DM}/\Omega_{\Lambda})\,(g\,{V_{\text{today}}}/{v})
\ll1$, which requires the sextic coupling to dominate the square of the quartic---a constraint on the microscopic theory drawn from a cosmological measurement.

{\bf Discussion.} --- In this Letter we have established a controlled analytical framework for interacting group field theory cosmology within single-mode dynamics in a simple but rather general class of models. For a model with quartic and sextic interactions, we derived the effective Friedmann equation in closed form for zero-mean Gaussian states, and expanded it in the small parameters that control the Gaussian closure to isolate reliable interaction effects. The cosmological impact of these interactions turns out to be rigidly constrained: they preserve the quantum bounce, dress the background parameters---including Newton's constant---with small corrections, and generate two new contributions. These behave effectively as a cosmological constant and a pressureless dark matter fluid, and emerge purely as collective quantum gravity effects. While remaining subdominant, they yield a closed-form expression for the ratio $\Omega_{\Lambda}/\Omega_{\rm DM}$, fixed by the fundamental theory at any given volume. 

The ratio \eqref{pheno} admits a simple reading: the currently observed $\Omega_{\Lambda}/\Omega_{\rm DM} \sim \mathcal{O}(1)$ places the present volume $V_{\text{today}}$ within a factor of order unity ($\approx 2.6$) beyond the GFT crossover scale $V_\star$. The value of $V_\star$ depends on the free parameter ratio $g/\lambda$ (up to initial-data corrections) and is not predicted \textit{a priori}, although consistency requires $g^{2}\ll\lambda$. At the same time, the proximity of $V_{\star}$ to $V_{\text{today}}$ provides a microscopic formulation of the cosmological coincidence problem~\cite{weinberg2008cosmology}, which translates into a dimensionless ratio of quantum gravity parameters.

Because the dark contributions remain subdominant to the scalar clock, the model constrains their relative importance rather than the dark sector's total share of the energy budget. What is established is therefore not a complete late-time cosmology, but that interactions  among fundamental ``quantum gravity constituents'' naturally generate energy densities possessing dust and cosmological-constant scalings, in a ratio fixed entirely by the fundamental dynamics. Whether these terms eventually dominate lies beyond the Gaussian closure and remains open. 

The correspondence established for quartic and sextic interactions reflects a general structural property: a $\hat\varphi^{2\kappa}$ interaction generates a leading effective fluid term with equation of state $w_{\kappa}=2-\kappa$, alongside dressings of lower-order scalings. The order of interaction alone dictates the characteristic fluid it effectively mimics (e.g., a phantom component for $\kappa=4$). In this sense, the dark sector is not an ad hoc input, but the first entries in a systematic dictionary mapping the GFT interaction order to an effective cosmological equation of state. 

The door is now open to several developments. Generic GFT interactions couple different Peter--Weyl modes; whether the dictionary survives beyond the single-mode reduction used here is a natural next question. The identification of the dust-like contribution with dark matter, in turn, rests on the background equation of state alone: confirming its physical nature requires analysing its perturbation dynamics beyond the homogeneous truncation. For displaced states, the dynamics can in principle be integrated numerically, but no autonomous Friedmann equation in the volume alone exists in terms of the conserved quantities identified here; this requires identifying additional invariants or applying other techniques. Finally, we note that the interactions leave the relative volume fluctuations unchanged: for any zero-mean Gaussian state $(\Delta V/V)^{2}=2+2 v/V-(Q-\tfrac14)v^2/V^{2}$, so they approach a constant as the universe grows---as in the free theory, where displacement is needed to suppress them \cite{Gauss}. Whether displacement operates similarly in the interacting theory remains an open question.

{\bf Acknowledgments.} --- 
DO and AC acknowledge support from Grant PR28/23 ATR2023-145735 funded by MCIN/AEI/10.13039/501100011033. The authors further acknowledge support from the WOST, WithOut SpaceTime project, supported by Grant ID 63683 from the John Templeton Foundation.

\bibliographystyle{apsrev4-2}
\bibliography{Refs}

\clearpage
\onecolumngrid 
\section*{Appendix}

In this appendix we derive the effective Friedmann equation \eqref{eq:ExactFE} and its expansion \eqref{Taylor}, establish the obstruction that motivates the restriction to vanishing displacement, and outline the systematic generalisation to arbitrary even-order interactions, making all underlying approximations explicit. 

\textbf{Interacting Hamiltonian and moment hierarchy.} --- We start from the Hamiltonian \eqref{Interacting_Hamiltonian}, namely
\begin{equation}\label{eq:SM_H}
\hat{H} = -\frac{\omega}{2}\big(\hat{a}^{2}+\hat{a}^{\dagger 2}\big) -\frac{g\omega}{4}\hat{X}^{4} -\frac{\lambda\omega}{6}\hat{X}^{6}\,,\quad\qquad g,\lambda>0\,,
\end{equation}
where we have defined $\hat{X}:=\hat{a}+\hat{a}^{\dagger}$, which corresponds to the group field $\hat \varphi$ up to a constant. We use $\alpha:=\langle \hat{a} \rangle\in\mathbb{C}$, $ k:=\langle \hat{a}^{2} \rangle\in\mathbb{C}$ and $n:=\langle \hat{N} \rangle\in\mathbb{R}_{\geq0}$, and we define for convenience the mean and the variance of the quadrature,
\begin{equation}\label{eq:SMb_xs}
x:=\langle \hat{X} \rangle=\alpha+\alpha^{*}\,,\qquad \sigma^{2}:=\langle \hat{X}^{2} \rangle-\langle \hat{X} \rangle^{2}=k + k^*+2n+1-x^{2}\,.
\end{equation}
Using $[\hat{a},\hat{X}^{m}]=m\hat{X}^{m-1}$ in the Heisenberg equation ${\rm i}\dd\hat{O}/\dd \chi=[\hat O,\hat{H}]$ for $\hat{a}$, $\hat{a}^{2}$ and $\hat{N}$ one finds the \textit{exact} identities
\begin{equation}\label{eq:SMb_heisenberg}
\begin{aligned}
    \hat{a}' &= {\rm i}\omega\big(\hat{a}^{\dagger}+g\hat{X}^{3}+\lambda\hat{X}^{5}\big)\,,\\
    (\hat{a}^{2})' &= {\rm i}\omega\big(2\hat{a}^{\dagger}\hat{a}+1\big) + {\rm i}g\omega\big(\hat{a}\hat{X}^{3}+\hat{X}^{3}\hat{a}\big) + {\rm i}\lambda\omega\big(\hat{a}\hat{X}^{5}+\hat{X}^{5}\hat{a}\big)\,,\\
    \hat{N}' &= -{\rm i}\omega\big(\hat{a}^{2}-\hat{a}^{\dagger 2}\big) + {\rm i}g\omega\big(\hat{a}^{\dagger}\hat{X}^{3}-\hat{X}^{3}\hat{a}\big) + {\rm i}\lambda\omega\big(\hat{a}^{\dagger}\hat{X}^{5}-\hat{X}^{5}\hat{a}\big)\,.
\end{aligned}
\end{equation}
Taking expectation values does not yield a closed system: the right-hand sides contain moments of order up to six, and their own equations involve even higher moments. In particular, commuting with a monomial of degree $d$ raises the maximal order by $d-2$; indeed, the quadratic part of $\hat{H}$ preserves it while $\hat{X}^4$ and $\hat{X}^6$ raise it by two and four. For $g,\lambda\neq0$ no finite set of moments closes on itself: the moment equations form an infinite hierarchy, equivalent to the full quantum dynamics---the moment analogue of the Bogolyubov--Born--Green--Kirkwood--Yvon (BBGKY) hierarchy of kinetic theory \cite{Bonitz2016}. The free theory yielding \eqref{Calcinari}
is solvable precisely because the order is preserved.

\textbf{Gaussian approximation.} --- The hierarchy is closed by the first of the approximations in the main text: we assume that the state is, and remains, Gaussian---fully specified by the five real parameters encoded in $\alpha$, $k$ and $n$. This is exact for quadratic Hamiltonians \cite{TextSerafini}, and an approximation otherwise: it discards the connected cumulants beyond second order, and is controlled by the conditions $gn\ll1$ and $\lambda n^{2}\ll1$ (see \eqref{gnl}). Gaussian states enable the use of Wick's theorem, whose precise statement requires the \textit{centred} operator $\hat{a}_{c}:=\hat{a}-\alpha$, its adjoint, and the corresponding quadrature $\hat{X}_{c}:=\hat{a}_{c}+\hat{a}_{c}^{\dagger}=\hat{X}-x$. For any ordered product $\hat Y_{1}\cdots\hat Y_{m}$ with $\hat Y_{i}\in\{\hat{a}_{c},\hat{a}_{c}^{\dagger}\}$,
\begin{equation}\label{eq:SMb_wick}
    \langle \hat Y_{1}\cdots\hat Y_{m} \rangle= \begin{cases} \displaystyle\sum_{P}\prod_{(i<j)\in P}\langle \hat Y_{i}\hat Y_{j} \rangle\,, & m\ \text{even}\,,\\
    0\,, & m\ \text{odd}\,, \end{cases}
\end{equation}
where the sum runs over the pairings $P$ of $\{1,\dots,m\}$. The only elementary contractions are
\begin{equation}\label{eq:SMb_contractions}
     k_c:=\langle \hat{a}_{c}^2 \rangle=k-\alpha^{2},\qquad n_c:=\langle \hat{a}_{c}^{\dagger}\hat{a}_{c} \rangle=n-|\alpha|^{2}\,,
\end{equation}
and of course $\langle \hat{a}_{c}^{\dagger\,2} \rangle=k_c^*$ and $ \langle \hat{a}_{c}\hat{a}_{c}^{\dagger} \rangle=n_c+1$, so that the variance is correspondingly $\sigma^{2}=\langle \hat{X}_{c}^{2} \rangle=k_c+k_c^*+2n_{c}+1$. 

Applying \eqref{eq:SMb_wick} to the structures appearing in \eqref{eq:SMb_heisenberg} yields the following identities. 
First, the Gaussian quadrature moments $X_{m}:=\langle \hat{X}^{m} \rangle$ can be expressed as
\begin{equation}\label{eq:SM_Xm}
    X_{m}=\sum_{j=0}^{\lfloor m/2\rfloor}\binom{m}{2j}(2j-1)!!\,\sigma^{2j}x^{m-2j}\,,
\end{equation}
satisfying $\partial_{x}X_{m}=mX_{m-1}$. 
For example, one computes
$X_{2}=x^{2}+\sigma^{2}$, $X_{3}=x^{3}+3x\sigma^{2}$, and so on. Note that, at zero mean, the odd ones
vanish and $X_{2j}=(2j-1)!!\sigma^{2j}$. Moreover, using $\hat{a}=\alpha+\hat{a}_{c}$ and $\hat{X}=x + \hat{X}_c$, one can write
\begin{equation}\label{eq:SM_Zm}
    Z_{m}:=\langle \hat{X}^{m}\hat{a} \rangle =\alpha X_{m}+\big(k_{c}+n_{c}\big)\partial_{x}X_{m}\,,
\end{equation}
where the coefficient $k_{c}+n_{c}=\langle\hat{X}_{c}\hat{a}_{c} \rangle$. Of course, the commutator fixes also $\langle \hat{a}\hat{X}^{m} \rangle=\langle \hat{X}^{m}\hat{a} \rangle+\langle [\hat{a},\hat{X}^{m}] \rangle=Z_{m}+\partial_{x}X_{m}$.
Using properties \eqref{eq:SM_Xm} and \eqref{eq:SM_Zm} in \eqref{eq:SMb_heisenberg} yields the closed system
\begin{equation}\label{eq:SM_five}
\begin{aligned}
     \alpha' &= {\rm i}\omega\big(\alpha^{*}+gX_{3}+\lambda X_{5}\big)\,,\\
     k' &= {\rm i}\omega\big[2n+1+g\big(2Z_{3}+\partial_{x}X_{3}\big) +\lambda\big(2Z_{5}+\partial_{x}X_{5}\big)\big]\,,\\
     n' &= {\rm i}\omega\big(k^{*}-k\big) + {\rm i}g\omega\big(Z_{3}^{*}-Z_{3}\big) + {\rm i}\lambda\omega\big(Z_{5}^{*}-Z_{5}\big) =2\omega\,\mathrm{Im}\big[k+gZ_{3}+\lambda Z_{5}\big]\,,
\end{aligned}
\end{equation}
with two complex and one real equation, matching the five real parameters of the Gaussian family. 

This general structure holds for \textit{any} even polynomial interaction
$U(\hat{X})=\sum_{\kappa=2}^{\kappa_{\rm max}}\frac{c_{\kappa}}{2\kappa}\hat{X}^{2\kappa}$ of highest power $2\kappa_{\rm max}$, which is the single-mode reduction of the interaction $U[\varphi]$ of \eqref{eq_action} in units of $\omega$. The Hamiltonian \eqref{eq:SM_H} is the case
$\kappa_{\rm max}=3$ with $(c_{2},c_{3})\equiv(g,\lambda)$. Writing $\hat{H}=-\frac{\omega}{2}\big(\hat{a}^{2}+\hat{a}^{\dagger 2}\big)-\omega U(\hat{X})$, the system takes the compact form
\begin{equation}\label{eq:SM_compact}
    \alpha'= {\rm i} \omega \big(\alpha^*+\langle U'\rangle\big)\,, \qquad k'={\rm i} \omega \big(2n+1+2\langle U'\hat{a}\rangle +\langle U''\rangle\big)\,, \qquad n'=2 \omega {\rm{Im}}\big[k+\langle U'\hat{a} \rangle\big]\,,
\end{equation}
where $\langle U' \rangle=\sum_{\kappa}c_{\kappa}X_{2\kappa-1}$, $\langle U'\hat{a} \rangle=\sum_{\kappa}c_{\kappa}Z_{2\kappa-1}$ and $\langle U'' \rangle=\partial_{x}\langle U' \rangle$. This shows that generalising to higher order interactions is straightforward; for example, including a $\hat{X}^{8}$ term requires adding $X_{7}$ and $Z_{7}$. In other words, higher interactions merely add terms to the sums in \eqref{eq:SM_compact}, without changing the structure. The system \eqref{eq:SM_five} is realised for $U(\hat{X})=\tfrac{g}{4}\hat{X}^4+\tfrac{\lambda}{6}\hat{X}^6$ so that  $\langle U'\rangle = g X_3 + \lambda X_5$ and
$\langle U'\hat{a} \rangle=gZ_{3}+\lambda Z_{5}=\alpha\langle U' \rangle+(k_{c}+n_{c})\langle U'' \rangle$. 

In the following, we will leverage the fact that for the system described by \eqref{eq:SM_five} there are two conserved quantities: the expectation value of the Hamiltonian and the symplectic invariant of the centred covariance matrix:
\begin{equation}\label{eq:SM_conserved}
    \langle \hat{H} \rangle=-{\omega}\,{\rm Re}\,k-\frac{g\omega}{4}X_{4}-\frac{\lambda\omega}{6}X_{6}\,, \qquad Q=\Big(n_{c}+\tfrac12\Big)^{2}-|k_{c}|^{2} \,,
\end{equation}
where $X_4$ and $X_6$ are computed from \eqref{eq:SM_Xm}, and $n_c$ and $k_c$ are given in \eqref{eq:SMb_contractions}. Notice that $\langle\hat{H}\rangle$ is conserved \textit{exactly}, because $\hat{H}$ is time independent and the Gaussian truncation affects only its specific expression. Conservation of $Q$, by contrast, is a property of the truncated flow itself, which acts on the centred covariance matrix by symplectic (Gaussian-unitary) transformations \cite{TextSerafini}.

\textbf{Obstruction from displacement.} --- In order to obtain an effective Friedmann equation like \eqref{eq:ExactFE}, we need a relation $(n')^{2}=f(n;\langle\hat{H}\rangle,Q)$, where $f$ is a function. Before restricting to the zero-displacement subspace adopted in the main text, it is natural to ask whether such a closure can hold for general, coherently displaced Gaussian states. Since $(n')^{2}$ is a property of a state, such an $f$ exists \textit{if and only if} every state sharing a triple $(n;\langle\hat{H}\rangle,Q)$ returns the same $(n')^{2}$.

Using $\langle U'\hat{a}\rangle=\alpha\langle U'\rangle+(k_{c}+n_{c})\langle U''\rangle$ and ${\rm Im}\,k={\rm Im}\,k_{c}+2{\rm Re}\,\alpha\,{\rm Im}\,\alpha$ in the last of \eqref{eq:SM_compact}, one finds
\begin{equation}\label{eq:SM_nprime_general}
\frac{n'}{2\omega}=\big(1+\langle U''\rangle\big)\,{\rm Im}\,k_{c}+\big(\langle U'\rangle+2{\rm Re}\,\alpha\big)\,{\rm Im}\,\alpha\,,
\end{equation}
where $\langle U'\rangle$ and $\langle U''\rangle$ depend on the state only through $x$ and
$\sigma^{2}$. Note that in the free-theory limit the right-hand side reduces to ${\rm Im}\,k$. Now fix $n$, $\langle\hat{H}\rangle$ and, for convenience, an arbitrary (non-zero) ${\rm Re}\,\alpha$. The expectation value of the Hamiltonian \eqref{eq:SM_conserved} then depends on the state only through ${\rm Re}\,k$, and strictly monotonically since $\partial\langle\hat{H}\rangle/\partial\,{\rm Re}\,k=-\omega(1+\langle U''\rangle)<0$. Thus, it fixes ${\rm Re}\,k$, and with it $\sigma^{2}$, $\langle U'\rangle$ and $\langle U''\rangle$. Only ${\rm Im}\,k_{c}$ and ${\rm Im}\,\alpha$ remain free, and the symplectic invariant $Q$ leaves the single condition
\begin{equation}\label{eq:SM_fibre}
\big({\rm Im}\,k_{c}\big)^{2}+\sigma^{2}\big({\rm Im}\,\alpha\big)^{2}=\frac{\sigma^{2}}{2}\Big(n+\frac12-{\rm Re}\,k\Big)-Q=:\mathcal{E}\,,
\end{equation}
where we used \eqref{eq:SM_conserved} with $n_{c}+\tfrac12+{\rm Re}\,k_{c}=\sigma^{2}/2$, and the right-hand side is a fixed number. In the $({\rm Im}\,k_{c},{\rm Im}\,\alpha)$-plane, \eqref{eq:SM_fibre} is an ellipse centred at the origin, with semi-axes $\sqrt{\mathcal{E}}$ and $\sqrt{\mathcal{E}}/\sigma$, whose points are precisely the states carrying the given triple. All such states are admissible, since the covariance matrix stays positive definite.

A closure would require $(n')^{2}$ to be constant on that ellipse, but \eqref{eq:SM_nprime_general} shows that $n'$ is a \textit{linear} function on the plane---so its level sets are parallel straight lines---and the one on which it vanishes passes through the origin and therefore meets the ellipse. A constant $(n')^{2}$ could then only be zero, meaning $n'$ would vanish on the whole ellipse---and a linear function does so only if it vanishes identically, which $1+\langle U''\rangle>0$ excludes. More explicitly, parametrising \eqref{eq:SM_fibre} by an angle $\theta$ defined as ${\rm Im}\,k_{c}=\sqrt{\mathcal{E}}\cos\theta$ and $\sigma\,{\rm Im}\,\alpha=\sqrt{\mathcal{E}}\sin\theta$,
\begin{equation}\label{eq:SM_angle}
\big(n'\big)^{2}=4\omega^{2}\mathcal{E}\left[\big(1+\langle U''\rangle\big)\cos\theta+\frac{\langle U'\rangle+2{\rm Re}\,\alpha}{\sigma}\sin\theta\right]^{2}\,,
\end{equation}
so that across states carrying identical conserved labels $(n')^{2}$ sweeps the whole interval from zero up to $4\omega^{2}\mathcal{E}\big[(1+\langle U''\rangle)^{2}+(\langle U'\rangle+2{\rm Re}\,\alpha)^{2}/\sigma^{2}\big]$. This shows that $f$ cannot exist since the squared bracket is constant only if the sine and cosine coefficients vanish, whereas here $1+\langle U''\rangle>0$ for $g,\lambda>0$.

Fixing ${\rm Re}\,\alpha$ was a convenience and not a restriction: a closure would have to hold on \emph{every} subset of the states carrying a given triple, so contradicting it on one suffices, and the value chosen is arbitrary. That it be non-zero is important since it guarantees that every state on the ellipse is displaced: for ${\rm Re}\,\alpha=0$ the ellipse would also contain the two zero-mean states ${\rm Im}\,\alpha=0$, and the argument would compare displaced with undisplaced states. The degenerate case $\mathcal{E}=0$ is harmless, forcing ${\rm Im}\,k_{c}={\rm Im}\,\alpha=0$ and hence $n'=0$, so that the ellipse collapses to a single state and there is nothing to compare. 

This excludes only relations based on $n$, $\langle\hat H\rangle$, and $Q$; it does not exclude closures using additional conserved quantities or restricted to lower-dimensional invariant submanifolds. Note that the free theory is special in this respect: for $g=\lambda=0$ the two terms of \eqref{eq:SM_nprime_general} recombine into $n'=2\omega\,{\rm Im}\,k$, so that the displacement drops out and $n'$ is fixed, once $n$ and $\langle\hat{H}\rangle$ are given, by the \emph{uncentred} invariant $(n+\tfrac12)^{2}-|k|^{2}$---which is no longer conserved in the interacting theory. That is why \eqref{Calcinari} holds in any state.

The obstruction therefore motivates restricting to a sector on which the dynamics closes, and that is preserved in evolution. The zero-displacement sector, $\alpha=0$, provides precisely such a reduction, and is the only linear condition on the displacement preserved by the flow. Indeed, $\alpha'={\rm i}\omega(\alpha^{*}+\langle U'\rangle)$ is purely imaginary when $\alpha$ is real, and purely real when $\alpha$ is imaginary, since then $x=0$ and $\langle U'\rangle$ vanishes, so both slices are left immediately. This sector owes its status precisely to the uncentred symplectic invariant, since for zero displacement $k_{c}=k$ and $n_{c}=n$, so that the quantity closing the free theory survives the interactions. Concretely, ${\rm Im}\,\alpha$ vanishes identically and \eqref{eq:SM_fibre} reduces to
$({\rm Im}\,k)^{2}=\mathcal{E}$: only the two states ${\rm Im}\,k=\pm\sqrt{\mathcal{E}}$ carry a given triple, and both return $(n')^{2}=4\omega^{2}\mathcal{E}\big(1+\langle U''\rangle\big)^{2}$.

We point out that in Ref.~\cite{Gielen_2020} a closed Friedmann-like equation was obtained---but only for the quartic theory and only at first order in the coupling---for Perelomov--Gilmore (i.e., zero-mean squeezed \cite{Gauss}) states: this was denoted ``remarkable'' there, and is naturally explained here since those states lie in the zero-mean sector. For Fock coherent states expressions were also given, but only for two chosen one-parameter families: coherent states sharing $\langle \hat{H}\rangle$ and $Q$ in general lead to different Friedmann-like equations and hence different cosmic histories. What our argument excludes is a relation of the form \eqref{eq:ExactFE} valid for arbitrary displaced states.

\textbf{Restriction to the zero-mean sector.} --- We therefore set $\alpha=0$, excluding coherent displacements. The restriction is consistent: $\hat{H}$ contains only even powers of $\hat{X}$, so $\alpha'$ in \eqref{eq:SM_five} involves
only $X_{3}$ and $X_{5}$, which vanish at $x=0$; hence $\alpha'=0$ holds identically on the sector defined by $\alpha=0$, and the subspace is preserved exactly. On this subspace, centred and full moments coincide, $n_{c}=n$ and $k_{c}=k$, and the variance reduces to
\begin{equation}\label{eq:SM_S}
S:=\sigma^{2}\big|_{x=0}=2{\rm Re}\,k+2n+1=\langle \hat{X}^{2} \rangle\,,\qquad \langle \hat{X}^{4} \rangle=3S^{2}\,,\qquad \langle \hat{X}^{6} \rangle=15S^{3}\,,
\end{equation}
so that $Z_{m}=(k+n)\,m\,X_{m-1}$, and in particular $Z_{3}=3(k+n)S$ and $Z_{5}=15(k+n)S^{2}$. The system \eqref{eq:SM_five} collapses to three real equations,
\begin{equation}\label{eq:SM_three}
 k'={\rm i}\omega\big[(2n+1)+(2k+2n+1)\big(3gS+15\lambda S^{2}\big)\big]\,,\qquad n'=2\omega\, {\rm Im}\,k\,\big(1+3gS+15\lambda S^{2}\big)\,.
\end{equation}
The factor $1+3gS+15\lambda S^{2}$ shows the enhancement due to interactions; the quartic and sextic terms enter \emph{additively}, a
consequence of the additivity of $\hat{H}$ and of the Wick contractions \eqref{eq:SM_Xm}. For a general even interaction it generalises to $1+\sum_{\kappa\geq2}(2\kappa-1)!!\,c_{\kappa}S^{\kappa-1}$. Now we must eliminate ${\rm Im}\,  k$ and ${\rm Re}\,  k$ (through $S$) from
\begin{equation}\label{eq:SM_star}
\left(\frac{n'}{n}\right)^{2}=4\omega^{2}\frac{(\mathrm{Im}\,k)^{2}}{n^{2}}\big[1+3gS+15\lambda S^{2}\big]^{2}\,,
\end{equation}
obtained by squaring the second equation in \eqref{eq:SM_three}. To do that, we use the two invariants \eqref{eq:SM_conserved}, which on this subspace read $Q=\big(n+\tfrac12\big)^{2}-|k|^{2}$ and $ \langle\hat{H}\rangle=-\omega\,\mathrm{Re}\,k-\tfrac{3g\omega}{4}S^{2}-\tfrac{5\lambda\omega}{2}S^{3}$, where we used $\langle \hat{X}^{2\kappa} \rangle=(2\kappa-1)!!\,S^{\kappa}$. Conservation of $Q$ fixes
$|k|^{2}=n(n+1)-n_{0}(n_{0}+1)+|k_{0}|^{2}$, where a subscript $0$ denotes initial values, so that we can write
\begin{equation}\label{eq:SM_B}
\big(\mathrm{Im}\,k\big)^{2}=n(n+1)-n_{0}(n_{0}+1)+|k_{0}|^{2}-\frac{1}{4}\left({S-2n-1}\right)^{2}\,,
\end{equation}
with $\mathrm{Re}\,k=\tfrac12\big(S-2n-1\big)$ from \eqref{eq:SM_S}. Similarly, the conservation of $\langle\hat{H}\rangle$ can be cast as a {cubic in $S$}:
\begin{equation}\label{eq:SM_cubic}
10\lambda S^{3}+3gS^{2}+2S=10\lambda S_{0}^{3}+3gS_{0}^{2}+2S_{0}+4(n-n_{0})=4n +2 -4 \langle\hat{H}\rangle/\omega\,, 
\end{equation}
which can be solved in closed form. The solution is unique: $S=\langle \hat{X}^{2} \rangle$ is positive, and since the left-hand side of
\eqref{eq:SM_cubic} vanishes at $S=0$ and grows monotonically for $S\geq0$ (its derivative is
$30\lambda S^{2}+6gS+2>0$ for $g,\lambda>0$), exactly one root is positive. Explicitly, setting
$t:=S+g/(10\lambda)$ removes the quadratic term and brings \eqref{eq:SM_cubic} to the depressed
form
\begin{equation}\label{eq:SM_depressed}
t^{3}+\mathcal{P}t=\mathcal{R}(n)\,,\qquad \mathcal{P}=\frac{20\lambda-3g^{2}}{100\lambda^{2}}\,,\qquad \mathcal{R}(n)=t_{0}^{3}+\mathcal{P}t_{0}+\frac{2(n-n_{0})}{5\lambda}\,,
\end{equation}
where $t_{0}=S_{0}+{g}/({10\lambda})$. Cardano's formula then gives
\begin{equation}\label{eq:SM_cardano}
S(n)=\sqrt[3]{\frac{\mathcal{R}}{2}+\sqrt{\Delta_{C}}}+\sqrt[3]{\frac{\mathcal{R}}{2}-\sqrt{\Delta_{C}}}-\frac{g}{10\lambda}\,,\qquad \Delta_{C}=\frac{\mathcal{R}^{2}}{4}+\frac{\mathcal{P}^{3}}{27}\,,
\end{equation}
the cube roots being taken so that $S(n)$ is the positive root of \eqref{eq:SM_cubic} (if $\Delta_{C}>0$ there is only one real root, and if $\Delta_{C}<0$ the two summands are complex conjugates and all three roots are real). The limits of vanishing coupling must be taken in \eqref{eq:SM_cubic}, not in
\eqref{eq:SM_cardano}: the shift $t$ and the coefficient $\mathcal{P}$ are singular as $\lambda\to0$,
whereas the cubic itself degenerates smoothly to the quartic-only and free cases.

\textbf{Effective Friedmann equation.} --- Finally, substituting \eqref{eq:SM_B} into \eqref{eq:SM_star}, with $S(n)$ given by \eqref{eq:SM_cardano}, yields the autonomous equation:
\begin{equation}\label{eq:SM_main}
\left(\frac{n'}{n}\right)^{2}=4\omega^{2}\big(1+3gS+15\lambda S^{2}\big)^{2}
\left[1+\frac{1}{n}-\frac{1}{n^{2}}\mathcal{I}_{0}
+\frac{1}{n^{2}}\Big((\mathrm{Re}\,k_{0})^{2}-(\mathrm{Re}\,k)^{2}\Big)\right],
\end{equation}
where $\mathrm{Re}\,k=\tfrac12(S-2n-1)$, and $\mathcal{I}_{0}=n_{0}(n_{0}+1)-(\mathrm{Im}\,k_{0})^{2}$ is the free-theory quantity defined in \eqref{Calcinari}. Defining $\mathcal{I}:=Q-\frac14+(\mathrm{Re}\,k)^{2}$ one can recast \eqref{eq:SM_main} as \eqref{eq:ExactFE}, using the volume observable $V=v\,n$. The square bracket is nothing but $(\mathrm{Im}\,k)^{2}/n^{2}$ (see \eqref{eq:SM_star}), and is what the symplectic invariant leaves once $S(n)$ is known. It is non-negative by construction and it vanishes exactly at the turning points, so it alone controls the bounce. On the other hand, the first bracket is the squared enhancement factor of \eqref{eq:SM_three}. This structure survives for higher interactions: only the enhancement
factor and the polynomial determining $S(n)$ change---an $\hat{X}^{8}$ term, for instance, adds $105c_{4}S^{3}$ to the former and turns the latter into a quartic. We remark that \eqref{eq:SM_main} is an exact algebraic identity following with no further approximation from the three-equation system \eqref{eq:SM_three}. For $g=\lambda=0$ the cubic \eqref{eq:SM_cubic} is linear,
$S=S_{0}+2(n-n_{0})$, so $\mathrm{Re}\,k=\mathrm{Re}\,k_0$, the enhancement factor is $1$, and \eqref{eq:SM_main} reduces precisely to the free-theory setting \eqref{Calcinari}.

Expanding $S(n)$ about its free value $S_{0}+2(n-n_{0})$ to first order in each coupling and
substituting in \eqref{eq:SM_main} yields
\begin{equation}\label{eq:SM_taylor}
\left(\frac{n'}{n}\right)^{2}=4\omega^{2}\Big[(1+g A_{0}+\lambda B_{0})
+\frac{1}{n}(1+gC_{0}+\lambda D_{0})
-\frac{1}{n^{2}}(\mathcal{I}_{0}+gE_{0}+\lambda F_{0})
+(12g+\lambda G_{0}){n}+120\lambda {n^{2}}+O(g^{2},\lambda^{2},g\lambda)\Big]\,,
\end{equation}
where the expansion is controlled by $gn$ and $\lambda n^{2}$, the same combinations that govern the validity of the Gaussian truncation \eqref{gnl}. Since $g$ enters always accompanied by one power of $n$ and $\lambda$ by two, a monomial $g^{p}\lambda^{q}$ contributes at leading order $n^{p+2q}$, so that expanding in the couplings is equivalent to expanding in $gn$ and $\lambda n^{2}$. The quartic and sextic corrections are moreover independent at this order because $g$ and $\lambda$ enter both \eqref{eq:SM_cubic} and the enhancement factor additively, cross terms first appearing at second order. The coefficients of \eqref{eq:SM_taylor} read
\begin{equation}\label{eq:SM_taylorcoeff}
\begin{aligned}
A_{0}&=9\big(k_{0}+k_{0}^{*}+2\big),\\
B_{0}&=30\big[(k_{0}+k_{0}^{*}+1)(2(k_{0}+k_{0}^{*})+5)-4\mathcal{I}_{0}\big],\\
C_{0}&=3\big[(k_{0}+k_{0}^{*}+1)(k_{0}+k_{0}^{*}+2)-4\mathcal{I}_{0}\big],\\
D_{0}&=15\big(k_{0}+k_{0}^{*}+1\big)\big[(k_{0}+k_{0}^{*}+1)(k_{0}+k_{0}^{*}+2)-8\mathcal{I}_{0}\big],\\
E_{0}&=3\big[2\mathcal{I}_{0}(k_{0}+k_{0}^{*}+1)+n_{0}(k_{0}+k_{0}^{*})(k_{0}+k_{0}^{*}+n_{0}+1)\big],\\
F_{0}&=\frac{5}{2}\big[12\mathcal{I}_{0}(k_{0}+k_{0}^{*}+1)^{2}+(k_{0}+k_{0}^{*})\big((k_{0}+k_{0}^{*}+2n_{0}+1)^{3}-(k_{0}+k_{0}^{*}+1)^{3}\big)\big],\\
G_{0}&=20\big[7(k_{0}+k_{0}^{*})+12\big]\,,
\end{aligned}
\end{equation}
where $k_{0}+k_{0}^{*}=2\mathrm{Re},k_{0}$, while the imaginary part of $k_{0}$ enters only through $\mathcal{I}_{0}$. Finally, constructing a Friedmann-like equation for the volume using $V=v\,n$, \eqref{eq:SM_taylor} takes the form of the main text, \eqref{Taylor}.

The structure of \eqref{eq:SM_taylor} persists for a general even interaction. Since at zeroth order in the couplings $S=2n+2{\rm Re}\,k_{0}+1$ exactly, expanding $S^{\kappa-1}$ shows that an $\hat{X}^{2\kappa}$ term contributes a leading $2\,(2\kappa-1)!!\,2^{\kappa-1}c_{\kappa}\,n^{\kappa-1}$, together with every power below it, down to the $n^{-2}$ term controlling the bounce. The coefficients of \eqref{eq:SM_taylorcoeff} are the dressings that the quartic and the sextic induce on all lower contributions. 
Only the leading coefficient is state-independent, while all lower ones carry the initial data. An $\hat{X}^{8}$ interaction would add $1680\,c_{4}n^{3}$ together with its own tower of dressings. This fixes the cosmological reading of the pattern. In terms of the relational clock $\chi$, a classical Friedmann equation sourced by perfect fluids of equations of state $w_{i}$ reads $({V'}/{V})^{2}=\sum_{i}a_{i}\,V^{\,1-w_{i}}$, with constants $a_{i}$ fixed by the present densities, since $\rho_{i}\propto V^{-(1+w_{i})}$ and $\pi_{\chi}$ is conserved. Each order therefore introduces one new effective equation of state, $w_{\kappa}=2-\kappa$: the kinetic term is the $\kappa=1$ member and returns the stiff clock, the quartic gives dust and the sextic a cosmological constant, while an $\hat{X}^{8}$ term would give a phantom fluid.

\end{document}